\documentstyle[11pt,aaspp4]{article}
\input psfig.tex
\begin{document}

\title{Ultra-high energy cosmic rays may come from clustered sources}
\author{John N. Bahcall}
\affil{Institute for Advanced Study, Princeton, New Jersey 08540}
\and
\author{Eli Waxman}
\affil{Department of Condensed-Matter Physics, Weizmann Institute,
Rehovot 76100, Israel}

\begin{abstract}
Clustering of cosmic-ray sources affects the flux observed beyond the
cutoff imposed by the cosmic microwave background and may be important
in interpreting the AGASA, Fly's Eye, and HiRes data.  The standard
deviation, $\sigma$, in the predicted number, $N$, of events above
$10^{20}$ eV is $\sigma /N = 0.9(r_0/10 {\rm \, Mpc})^{0.9}$, where
$r_0$ is the unknown scale length of the correlation function ($r_0
\simeq 10$ Mpc for field galaxies, $H_0 = 50$ km s$^{-1}$ Mpc$^{-1}$).
Future experiments will allow the determination of $r_0$ through the
detection of anisotropies in arrival directions of $\sim10^{20}$~eV
cosmic-rays over angular scales of $\Theta\sim r_0/30$~Mpc.

\end{abstract}

\section{Introduction}

The conventional astronomical picture for the origin of ultra-high
energy cosmic rays, namely proton acceleration to high energy in
extra-galactic objects, predicts a sharp suppression of the cosmic-ray
flux beyond the Greissen-Zatsepin-Kuzmin (GZK) cutoff at $\sim 5
\times 10^{19}$ eV (Greisen 1966; Zatsepin \& Kuzmin 1996), due to
interaction of protons with photons of the cosmic microwave
background.  The absence of a GZK cutoff might suggest the presence of
a new source of ultra-high energy cosmic rays, possibly related to the
decay of exotic particles (see Cronin 1996; Hillas 1998 for reviews).

The suppression of flux beyond the GZK cutoff is most often discussed
assuming a uniform source distribution. However, the distribution of
other astrophysical systems (e.g. galaxies, clusters of galaxies) is
inhomogeneous on scales of tens of Mpc, comparable to the propagation
distance of protons of energy $>5\times10^{19}$~eV.  Therefore,
significant deviations from the predictions for a uniform distribution
may be expected beyond the GZK cutoff (Giller, Wdowczyk \& Wolfendale
1980; Hill \& Schramm 1985; Waxman 1995).

There are at least two possible approaches to evaluating the effects
of source inhomogeneity on the energy spectrum and spatial direction of
high energy cosmic rays. In the first approach, one assumes that the
source density of ultra-high energy cosmic rays is proportional to
(possibly with some bias factor) the galaxy density in some particular
survey of the distribution of relatively nearby galaxies. This
approach has been used by Waxman, Fisher \& Piran (1997); Giller,
Wdowczyk \& Wolfendale (1980);and Hill \& Schramm (1985) and
illustrates some of the principal features of source clustering. In the
present paper, we adopt a different and complimentary approach. We use
an analytic model that summarizes the clustering properties of the
unknown source of ultra-high-energy cosmic rays by a single parameter,
$r_0$, the correlation length.

The analytic model that we adopt has
the advantage of generality. No specific source population has to be
assumed. Also, the results of source clustering on the energy spectrum
and on the spatial distribution of the highest energy cosmic rays can
be summarized in terms of the single unknown correlation length of the
source population, $r_0$.  Thus the  correlation length provides a
concise and simple characterization of measurable
clustering effects on observational parameters for high energy cosmic
rays. Future experiments that measure the arrival directions of a large
number of high energy cosmic rays will determine $r_0$ and permit a clear
assessment of whether or not the departures from expectations based
upon a homogeneous source distribution can plausibly be explained by
any non-uniform population.

The source correlation function, $\xi({\rm d}\vec{r})$, is defined by
an average over the observable universe volume, $\langle
n(\vec{r})n(\vec{r}+{\rm d}\vec{r})\rangle\equiv \langle
n\rangle^2(1+\xi({\rm d}\vec{r}))$ where $n$ is the source density and
brackets denote volume average. Thus, the variance that we derive in
the expected cosmic-ray number due to source clustering is, strictly
speaking, the variance of the distribution of cosmic-ray number
observed by observers randomly distributed over the universe.  This
is the best that one can do, but it is not exactly what we we want. We
really want the variance in conceivable realizations of the universe
in our vicinity.  However, we show in \S2 that the distance from our
Galaxy over which sources contribute to the observed flux above
$10^{20}$~eV is $\sim40$~Mpc (see Fig. 1). Since the clustering
properties of astronomical objects (e.g. galaxies) within this volume
are similar to that obtained by averaging over larger volumes
(e.g. Peebles 1999), we expect our result to provide a reasonable
approximation for the variance in the number of cosmic-rays observed
at Earth due to different realizations of cosmic-ray source
distributions that satisfy the universal clustering properties of the
sources.

One may argue that a more accurate estimate of the variance
can be derived by the using known galaxy catalogs to model the effects of
source inhomogeneity. However, in order to use such an approach one
must choose a specific model to derive the probability of a given
cosmic-ray source distribution under the constraints provided by
a given galaxy catalog. The variance calculated in this approach would depend
primarily on the resultant cosmic-ray source correlation function. While
this correlation function would indeed depend on the clustering properties
of the particular galaxies chosen, it would be strongly dependent on
the assumed model. It should be emphasized here that one
does not known which particular rare sub-population of the total
observed galaxy population actually produces the highest energy cosmic
rays. The local density of field galaxies is of order $0.1~{\rm
Mpc^{-3}}$, whereas the local density of sources of UHE cosmic rays
may be as low as $0.00002~{\rm Mpc^{-3}}$.
(The lower limit to the source density is set by
the requirement that at least a few sources exist out to a distance of
$\sim 40 ~{\rm Mpc}$ in order to account for the observed events above
$10^{20} ~{\rm eV}$.)
Thus, results derived in this approach would not
yield a more accurate estimate of the variance, but rather reflect the
assumptions under which the cosmic-ray source distribution is constrained
given some chosen galaxy catalog. Moreover, even
after one makes a specific assumption about the functional relation
between the density of sources of ultra-high energy cosmic rays and an
observed sample of galaxies, one has to perform and analyze many Monte
Carlo simulations with galaxy catalogs in order to determine the
statistical effects on the observed characteristics of ultra-high
energy cosmic rays.

We believe therefore that the simplicity and conciseness of the
analytic model approach justifies its application, as one of the possible
approaches, to the evaluation of the effects of source inhomogeneity.

In this paper we consider the implications of clustering of
cosmic-rays sources, adopting the conventional picture of protons as
the ultra-high energy(UHE) cosmic-rays.  We assume that the
correlation function between the sources of ultra-high energy cosmic
rays has the same functional form as for galaxies, clusters of
galaxies, and quasars, but we do not require---as was done in previous
work (Waxman, Fisher \& Piran 1997; Giller, Wdowczyk \& Wolfendale
1980; Hill \& Schramm 1985)---that the distribution of cosmic ray
sources be the same as for nearby galaxies.  As described above, the
conditions that must be satisfied in order to produce ultra-high
energy cosmic rays are so exceptional that the spatial distribution of
sources of UHE cosmic rays could be very different than the distribution
of average, nearby galaxies like the IRAS sources discussed in Waxman,
Fisher \& Piran (1997). Also, we are primarily concerned in this paper with
the energy spectrum of UHE cosmic rays, whereas the IRAS galaxy
distribution was used in Waxman, Fisher \& Piran (1997) to discuss
their possible angular distribution on the sky.

Our principal result is that the standard deviation due to clustering,
$\sigma_{\rm clustering}(E_c)$, in the number of cosmic rays detected
above energy $E_c$ is proportional to the number, $N_{\rm
smooth}(E_c)$, predicted for a uniform source density.  This behavior
contrasts with the standard deviation due to shot noise, which
becomes proportional to $N_{\rm smooth}^{1/2}$. The constant of
proportionality between $\sigma_{\rm clustering}$ and $N_{\rm smooth}$
depends upon energy and is calculated in Section 2.  For
$E_c\sim10^{20}$~eV, the constant of proportionality between
$\sigma_{\rm clustering}$ and $N_{\rm smooth}$ is of order unity for
plausible values of the correlation length, $r_0$, of the correlation
function of the sources of the UHE cosmic rays.  Anisotropies 
in the  source distribution
on a scale of $\Theta \sim \left(r_0/100{\rm\,Mpc}\right)$
should be detectable 
in the angular distribution of $\sim10^{19.7}$~eV cosmic rays
with the high
rates that will be observed in the HiRes (Corbato et al. 1992), the
Auger (Cronin 1992; Watson 1993) and the Telescope Array (Teshima
1992) experiments.  Unlike the predictions of many particle-physics
explanations (for recent reviews see Berezinsky 1998; Bhattacharjee
1998) of the Fly's Eye (Bird et al. 1993, 1994) and AGASA (Hayashida
et al. 1994; Yoshida et al. 1995; Takeda et al. 1998) results, no
characteristic dependence on Galactic coordinates is expected on the
basis of this conventional extragalactic scenario.

The computed large value of $\sigma_{\rm clustering}$ reflects
the fact that the universe, and hence the UHE source population, is
inhomogeneous over distances comparable to the mean free path of UHE
protons that move through the cosmic microwave background radiation.
The ratio $\sigma_{\rm clustering} /N_{\rm smooth}$ cannot be reduced
by observing longer or doing different experiments. We only have one
``nearby universe'' and we do not know {\it a priori} the local
clustering properties of the sources of UHE cosmic rays.

The angular scale of anisotropies due to magnetic scattering of UHE
cosmic rays from individual sources is smaller than what is expected
from the clustering of the sources themselves unless the
inter-galactic field is close to the maximum value allowed with the
available data on Faraday rotation. The upper limit on the
contribution of an inter-galactic magentic field to the Farady
rotation of distant sources, RM$<1{\rm rad/m}^2$ for sources at
$z=2.5$ (Vall\'ee 1990), implies an upper limit $B<10^{-10}$~G on a
uniform inter-galactic field, and an upper limit
$B\lambda^{1/2}<10^{-8}{\rm G\,Mpc}^{1/2}$ on a field with correlation
length $\lambda$.\footnote{ The value we quote is a few times larger
than quoted by Vall\'ee, since the upper limit is inversly propotional
to the electron density $n_e$ and we choose $n_e=3\times10^{-7}{\rm
cm}^{-3}$, corresponding to $\Omega_b h^2=0.03$, while Vall\'ee
assumed $n_e=10^{-6}{\rm cm}^{-3}$.  Our upper limit is much smaller
than that recently claimed by Farrar \& Piran (2000), mainly due to
the fact that they use older radio data (Kronberg \& Simard-Normandin
1976) for which the RM upper limit for $z=2.5$ sources is weaker,
RM$<5{\rm rad/m}^2$, and assume that the free electron density is only
$0.3$ of the baryon density.} A simple random walk calculation shows
that the upper limit on $B\lambda^{1/2}$ sets an upper limit to the
magnetic deflections, $\theta_s<0.04(d/1{\rm Mpc})^{1/2}(E/10^{20}{\rm
eV})^{-1}$ for propagation distance $d$.\footnote{Following the
suggestion of Kulsrud {\it et al.} (1997), that magnetic fields could
be amplified by turbulence associated with the formation of large
scale filaments and sheets, several authors have recently considered
propagation of cosmic-rays in large scale inter-galactic magnetic
field (e.g. Sigl {\it et al.} 1999, Farrar \& Piran 2000). The upper
limit on magnetic field coherent on 1~Mpc scale in such structures is
smaller by a factor $f^{1/2}$ than the upper limit of $10^{-8}$~G on a
volume filling field, where $f$ is the fraction of the volume occupied
by large scale filaments and sheets. This is due to the fact that the
upper limit is inversly proportional to the electron density,
$n_e\propto f^{-1}$, and propotional to the square root of the path
length $l\propto f$ of light through magnetized plasma. While this
reduction of the upper limit may be partly compensated by assuming
strong negative evolution of the magnetic field with redhsift
(e.g. Farrar \& Piran 2000), the assumption that the inter-galactic
magnetic field is confined to large-scale structires does not lead to
increase in the upper limit on the deflection of UHE cosmic rays.}
Galactic magnetic fields contribute only relatively small deflections.

We present our principal results in Section~\ref{sec:results}.
We show in Section~\ref{sec:discussion} that the large value of
$\sigma_{\rm clustering} /N_{\rm smooth}$  is
important for the interpretation of the existing AGASA and Fly's Eye
experiments. We also list in Section~\ref{sec:discussion} the effects
of clustering that may be detectable with the HiRes, Auger project,
and the Telescope Array experiment.

\section{Results}
\label{sec:results}

The number, $N$, of high-energy cosmic rays predicted on average
 above a threshold
energy $E_c$ is
\begin{eqnarray}
N_{\rm smooth}(E_c) ~\equiv~ \langle N\left(E\geq E_c\right)\rangle &=& \int^\infty_{E_c}dE \int d^3
\vec{r} P \left(\vec{r}, E; E_c\right)\cr
&&\langle\left(\partial S/\partial
E\right)_{\vec{r}}n\left(\vec{r}\right)\rangle/4\pi r^2  ,
\label{ndefinition}
\end{eqnarray}
where $P \left(\vec{r}, E; E_c\right)$ is the probability that a
proton created at $\vec{r}$ with energy $E$ arrives at earth with an
energy above threshold.
The
energy spectrum generated by the sources is $\partial S/\partial
E$ and the luminosity-weighted source density is $ n\left(\vec{r}\right)$.
We present calculations for
$\partial S/\partial E \propto E^{-2}$ and $\partial S/\partial
E \propto E^{-3}$, which spans the range usually considered in the literature.
The AGASA and Fly's Eye results above $10^{19}$ eV can be shown to
require a spectrum less steep than $E^{-2.8}$ (Waxman 1995).

It is convenient to rewrite Eq.~(\ref{ndefinition}) in terms of a
function $F\left(r\right)$, where

\begin{equation}
\langle N\left(E\geq E_c\right)\rangle = \int^\infty_{0} dr
F\left( r \right)
\langle n \rangle ,
\label{nintermsofF}
\end{equation}
and $F$ is the survival probability averaged over energies,
\begin{equation}
F(r) \equiv \int^\infty_{E_c} dE P\left(r, E; E_c\right)\left(\partial
S/\partial E\right) .
\label{fdefinition}
\end{equation}
The constant of proportionality between  $\partial S/\partial E $
and $E^{-m}$ can be chosen arbitrarily since the constant cancels
out of the ratio of standard deviation to expected number, which is
the primary quantity we calculate (see Eq.~\ref{sigmaoverngeneral}).
For convenience, we fix the constant
so that $F(0) = 1.0$; we also suppress the dependence of $F$ on $E_c$.

For ultra-high energy cosmic rays ($E > 5\times 10^{19}$ eV),
the probability $P(r,E;E_c)$ is determined by the energy loss of protons
due to pair and pion production in interaction with the microwave background.
The energy
loss in a single pair production interaction is small, of order $m_e/m_p$,
and for protons at energy $>10^{19}$~eV the characteristic energy loss time
due to this interaction is comparable to the Hubble time,
$\approx5\times10^9$~yr (Blumenthal 1970). Hence, pair production
has only a small effect on $P(r,E;E_c)$ for $E_c\ge10^{20}$~eV,
where the proton life time due to pion production is $\le3\times10^{8}$~yr.
We  calculated therefore $P(r,E;E_c)$ taking into account
pion production interactions only. We  used
the compilation of cross sections given in Hikasa et al. (1992),
and assumed that the total cross section corresponds to the isotropic
production of a single pion (this assumption has been shown to be
adequate in Yoshida \& Teshima 1993; Aharonian \& Cronin 1994). The resulting probability
distribution may be approximated well over the energy range of interest
by a function of the form
\begin{equation}
P\left(r, E; E_c\right) = \exp\left[- a\left(E_c\right) r^2 \exp
\left(b\left(E_c\right)/E\right)\right].
\label{pgeneraldefinition}
\end{equation}
We give below results obtained using a precise  numerical evaluation,
of $P(r,E;E_c)$. The approximation (\ref{pgeneraldefinition})
provides a simple description of the functional dependence
of $P$ on $r$ and $E$ and can be used above $8\times 10^{19}$ eV to
evaluate to an accuracy of $10$\% the
average quantities considered in this paper.
The appropriate values of $a$ and $b$ for $E_c/ (10^{20}~{\rm eV})
=1, 3, {\rm ~and ~} 6$  are, respectively,
$a/(10^{-4}{\rm Mpc}^{-2}) = 1.4, 9.2, {\rm and~} 11 $,
$b/(10^{20}{\rm ~eV}) = 2.4, 12, {\rm  and~} 28$.

The function $F(r)$ is shown in Fig. 1 for a spectrum
$\partial S/\partial E\propto E^{-2}$ .
The average distance from which a proton originates
if it is observed at earth to have an energy in excess of $E_c$,
\begin{equation}
\langle r \rangle \equiv {{\int dr r F(r) } \over  {\int dr F(r)} },
\label{average definition}
\end{equation}
can be calculated  analytically using Eq.~(\ref{fdefinition}) and
Eq.~(\ref{pgeneraldefinition}).  For an energy spectrum with
$\partial S/\partial E \propto E^{-2}$,

\begin{equation}
\langle r \rangle =
\frac{\left[ 1 - \exp(-b/E_c)\right]}
{\sqrt{4 \pi a}\left[1 - \exp(-b/2E_c)\right]} .
\label{averagedistance}
\end{equation}
A similar expression can be obtained for
$\partial S/\partial E \propto E^{-3}$.
Inserting the appropriate values of $a$ and $b$
in Eq.~(\ref{averagedistance}),
$\langle r \rangle = 31.2$ Mpc for $E_c = 1\times 10^{20}$ eV and
$\langle r \rangle = 10.6$ Mpc for $E_c = 3\times 10^{20}$ eV (the
values  obtained using the  numerical calculation of $F(r)$ are
$31.8$~Mpc and $10.9$~Mpc, respectively).
The observed universe is inhomogeneous on these distance scales.

The variance in the number of cosmic rays observed above an energy
$E_c$ can be computed from the expression
\begin{equation}
\sigma^2\left(E \geq E_c\right) = \langle N^2\left(E \geq
E_c\right)\rangle - \langle N\left(E \geq E_c\right)\rangle^2 .
\label{sigmadefinition}
\end{equation}
Since galaxies, clusters of galaxies, and quasars are all clustered
with correlation functions that have the same shape, it is natural to
suppose that the sources of UHE cosmic rays are also
clustered with a similar correlation function. Explicitly, we assume
that
\begin{equation}
\langle n\left(\vec{r}^{\,\prime}\right)
n\left(\vec{r}^{\,\prime\prime}\right)\rangle \equiv
\langle n \rangle^2
\left(1 + \xi
\left(\vec{r}^{\,\prime} - \vec{r}^{\,\prime\prime}\right)
+ \delta(\vec{r}^{\, \prime} -\vec{r}^{\,\prime\prime})/\langle n \rangle 
\right) ,
\label{correlationdefine}
\end{equation}
where the correlation function $\xi$ is

\begin{equation}
\xi (r) \equiv \left(\frac{r_0}{r}\right)^{1.8} 
\label{zetadefinition}
\end{equation}
and the $\delta$ function contribution represents shot-noise.
The correlation length, $r_0$, is not known for the sources of
UHE cosmic rays, but it is about $10$ Mpc for galaxies (Groth \&
Peebles 1977. Shectman et al. 1996, Tucker, Oemler, Kirshner et al. 1997)
and about $40$ Mpc for rich clusters of galaxies (Bahcall 1985;
Bahcall \& Soneira 1983; Peacock and West 1992)
(for $H_0  = 50$
km s$^{-1}$ Mpc$^{-1}$).

When the variance is calculated using Eq.~(\ref{sigmadefinition}),
the only term that
survives, in addition to shot noise, is
proportional to the correlation function, i.e.,
\begin{eqnarray}
\sigma^2
\left(E \right. & \geq & \left. E_c\right) =  \int\int dE^\prime
dE^{\prime\prime}\left(\partial S/\partial E^\prime\right)
\left(\partial S/\partial E^{\prime\prime}\right)\nonumber\\
&& \int\int  \frac{d^3 \vec{r}^{\,\prime}}{4\pi{r^\prime}^2} \frac{d^3
\vec{r}^{\,\prime\prime}}{4\pi {r^{\prime\prime}}^2}
P\left(\vec{r}^{\,\prime}, E^\prime; E_c\right)
 P\left(\vec{r}^{\,\prime\prime}, E^{\prime\prime}; E_c\right)\nonumber\\
&\times&
\langle n\left(\vec{r}^{\,\prime}\right)\rangle \langle
n\left(\vec{r}^{\,\prime\prime}\right)\rangle \xi \left(\vec{r}^{\,\prime}
- \vec{r}^{\,\prime\prime}\right) ~+~ \sigma^2_{\rm shot-noise}.
\label{sigmageneral}
\end{eqnarray}
For simplicity, we will not display
 $\sigma_{\rm shot-noise}$ explicitly in what follows.
After some algebra, we find

\begin{eqnarray}
&&\frac{\sigma_{\rm clustering}}{ N_{\rm smooth}} =
\left(5/2\right)^{1/2} r^{0.9}_0\nonumber\\
&&\frac{\left[\int^\infty_0\int^\infty_0
\frac{dxdy}{xy}F(x)F(y)\left\{(x + y)^{0.2} - \bigl|(x -
y)\bigr|^{0.2}\right\}\right]^{0.5}}{\int^\infty_0 dr F(r)} .
\label{sigmaoverngeneral}
\end{eqnarray}

The ratio $
{\sigma_{\rm clustering}}
/{ N_{\rm smooth}}$ can be evaluated numerically using
Eq.~(\ref{fdefinition}) and  Eq.~(\ref{sigmaoverngeneral}).
For cosmic rays above $1.0\times 10^{20}$ eV, we find

\begin{equation}
\frac{\sigma_{\rm clustering}}{ N_{\rm smooth}}
= 0.9 \left({r_0 \over {10~{\rm Mpc}}}
\right)^{0.9} ,
\label{sigma1*20}
\end{equation}
and for cosmic rays above $3.0\times 10^{20}$ eV,

\begin{equation}
\frac{\sigma_{\rm clustering}}{ N_{\rm smooth}} = 2.6 \left({r_0 \over {10~{\rm Mpc}}}
\right)^{0.9} .
\label{sigma3*20}
\end{equation}
The results given in Eq.~(\ref{sigma1*20}) and Eq.~(\ref{sigma3*20})
were calculated assuming a spectrum $\propto E^{-2}$. If the spectrum
is instead assumed to be
$\propto E^{-3}$, then the coefficient $0.9$ in Eq.~(\ref{sigma1*20})
is increased  to
$1.0$ and the coefficient in Eq.~(\ref{sigma3*20}) becomes $3.6$.
For cosmic rays above $3.0\times 10^{20}$ eV, the dispersion
is so large (see Eq.~\ref{sigma3*20} ) that it will be
difficult to interpret statistically  the
observed number.

Due to the rapid decrease with $r$ of $F(r)$ at large $r$, shown in
Fig. (1), the integral (\ref{sigmageneral}) is dominated by the
contribution from small separations $|\vec r'-\vec r''|$. Neglecting, for
example,
the contribution from large separations $|\vec r'-\vec r''|>20$~Mpc,
where the galaxy-galaxy correlation function drops faster than
$r^{-1.8}$, reduces $\sigma_{\rm clustering}
/{ N_{\rm smooth}}$ by 20\% for $E_c=10^{20}$~eV
and by 3\% for $E_c=3\times10^{20}$~eV. Thus  our results for
$\sigma_{\rm clustering}/{ N_{\rm smooth}}$
are not sensitive to the $r$ dependence of $\xi(r)$ at large $r$.
On the other hand, $\sigma_{\rm clustering}
/{ N_{\rm smooth}}$ is more sensitive to the $r$
dependence at small $r$. For $\xi(r)\propto r^{-1.8}$ and $r_0 = 10$ Mpc,
approximately half the contribution to the integral (\ref{sigmageneral}) comes
from separations $|\vec r'-\vec r''|<3$~Mpc for
$E_c=10^{20}$eV and $|\vec r'-\vec r''|<1$~Mpc for $E_c=3\times10^{20}$~eV.
The galaxy-galaxy correlation
function, for example, is known to follow the form (\ref{zetadefinition})
down to $r\sim0.1$~Mpc, while the cluster-cluster correlation function is not
well measured for $r<10$~Mpc.

In principle, the contribution to the variance from shot-noise,
 $\sigma_{\rm shot-noise}^2$, could be large 
due to the possibility of having a
 nearby source. However, an extremely close source would dominate the
 all sky flux, which is not the case for the observed cosmic-ray
 events.  If the average source density is much larger than the
 minimum value required to explain the observations of UHE cosmic
 rays, i.e. $<n> ~ >>~ 10^{-4}~{\rm Mpc^{-3}}$, then the shot noise
 contribution is relatively small. The shot noise could be
 significant, of order $N_{\rm smooth}$ , if the local source density
 is of order $10^{-4}~{\rm Mpc^{-3}}$.

\section{Discussion and Predictions}
\label{sec:discussion}

Are  new particles required to explain the
observed number of ultra-high energy cosmic rays?
The  number of events observed beyond $10^{20}$ eV by AGASA
 (Takeda et al. 1998),  by the Fly's Eye (Bird et al. 1993, 1994), and
by Yakutsk (Efimov et al. 1991) is, respectively,
$6$, $1$,  and $1$.
The average observed number of events, $2.7$,  is to be compared with
conventional
estimates of the  number of events from a smooth model of between
$N_{\rm smooth} = 0.7$
(for $\partial S/\partial E \propto E^{-3}$) (cf. Takeda et al. 1998)
and $N_{\rm smooth} = 2.2$(for $\partial S/\partial E \propto E^{-2}$)
(cf. Waxman 1995).
The total standard deviation in the expected number of events beyond
$10^{20}$ eV is
\begin{equation}
\sigma_{\rm total} = \left[0.8\left(r_0/10 ~{\rm Mpc}\right)^{1.8}
N^2_{\rm smooth} + N_{\rm smooth} \right]^{0.5},
\label{totalsigma}
\end{equation}
where the first term in the brackets is the calculated uncertainty due to
clustering  and the second term is  due to Poisson statistics.
For $N_{\rm smooth} = 2.2$ and $r_0 = 10$ Mpc,  $\sigma_{\rm total} = 2.4$.
Even for $N_{\rm smooth} = 0.7$, $\sigma_{\rm total} = 1.0$.
No matter how one combines the experiments and the smooth models,
there is not a $3\sigma$
discrepancy with the conventional  picture of protons as the source of
ultra-high energy cosmic rays.
If one considers, as many recent authors have done, only the AGASA
data, then $N_{\rm observed} = 6$ and
for $\partial S/\partial E \propto E^{-2}$, $N_{\rm smooth}
= 2.2$, and $\sigma_{\rm total} = 2.4$, which is a $1.6\sigma$
discrepancy.
On the basis of the available  evidence, 
we conclude that new particles are
not required to explain the observed cosmic ray energy spectrum.

However, a number of authors have interpreted the same data as
suggesting the possibility that new particles are producing the
highest energy cosmic rays (for a recent discussion and review of this
point of view see Ellis 1999 and the references contained
therein). One of the principal reasons for the difference in
conclusions is the uncertainty over whether the extrapolation of the
`conventional high energy component' should be made using $\partial
S/\partial E \propto E^{-2}$, as we have done, or whether the fall-off
of the spectrum at energies just below the GZK cutoff is stronger,
e. g., $\propto E^{-3}$. This is just one of the important questions
that will have to be answered in the future by a much larger data set.

We also note that preliminary results from the HiRes experiment were
recently presented in a talk at TAUP99 (Matthews 1999), reporting 7
events beyond $10^{20}$~eV for an exposure similar to that of the
Fly's Eye.  It is difficult to decide how this result should be
interpreted, since the discrepancy between HiRes and Fly's Eye results
is present not only above $10^{20}$~eV, but also at lower energy,
where Fly's Eye, AGASA and Yakutsk experiments are in agreement: 13
events above $6\times10^{19}$~eV are reported in the preliminary HiRes
analysis, while only 5 events at that energy range are reported by
Fly's Eye. We therefore believe that unambiguous conclusions based on
the recent HiRes data can be drawn only after a complete analysis of
the HiRes data is published (which would include, e.g., corrections
due to realistic atmospheric conditions).

One possible interpretation of the suggested excess of events beyond
the GZK cutoff is that this may be a hint that ultra-high energy
cosmic rays are produced by sources whose density in the nearby
universe is somewhat higher than their average cosmological density.
Of course, other interpretations are possible, including the
hypothesis--amply represented in the literature--that new physics is
involved and that the GZK cutoff is bypassed by this new physics. If
the GZK cutoff is ultimately convincingly measured, then the most
likely explanation for the small number of already observed UHE cosmic
rays will be the effects of source clustering.

Future measurements of high statistical significance with
AGASA, HiRes, Auger and the Telescope Array
are required to determine if source  clustering affects
significantly the observed number
and  angular distribution of UHE cosmic rays.
The predicted   effects of clustering depend upon
the unknown correlation length $r_0$.

We summarize below the principal observational implications of
assuming a correlation length between $10$ Mpc (observed for galaxies)
and $40$ Mpc (observed for the rich clusters of galaxies).

\noindent
1) Clustering gives rise to a standard deviation
 that is proportional to $N_{\rm smooth}$ (see
Eq.~(\ref{sigma1*20}) and Eq.~(\ref{sigma3*20}) ).
When more events beyond the GZK cutoff are available,
clustering will be more important than shot noise
in determining
the shape of the energy spectrum of the highest energy cosmic rays.
Even for the existing Fly's Eye and AGASA samples, the
 variance due to clustering may dominate shot noise (see
 Eq.~(\ref{totalsigma})).

\noindent
2) The fractional calculated variance, $\sigma_{\rm clustering} /{
N_{\rm smooth}}$, increases with the cutoff energy, $E_c$, which
simply reflects the fact that the mean distance from which protons
originate decreases with the observed energy (see
Eq~\ref{averagedistance}). The fractional variance is large at all
energies beyond the cutoff. With the approximations of the present
paper, $\langle r \rangle\simeq120$~Mpc and $\sigma_{\rm clustering}
/{ N_{\rm smooth}} = 0.25 \left({r_0 / {10~{\rm Mpc}}} \right)^{0.9}$
for $E_c=6\times10^{19}$~eV [the exact value of $\sigma/{ N_{\rm
smooth}}$ ($\langle r \rangle$) is somewhat larger since at this
energy pair production energy loss, neglected in our calculations,
becomes important]. The fractional variance does not increase
significantly as $E_c$ is increased above $3\times10^{20}$~eV, since
the average distance at which a proton originates is approximately
constant at higher energy (e.g., $\langle r \rangle=9.5$~Mpc for
$E_c=6\times10^{20}$~eV, comparable to $\langle r \rangle=10.9$~Mpc
for $E_c=3\times10^{20}$~eV).

\noindent 3)
Anisotropies should be observed on large angular
scales $\Theta_0 \sim r_0/r_{\rm origin}$,
where $r_{\rm origin}$ decreases from $\sim 200$ Mpc to $\sim 30$ Mpc
as $E_c$ increases from $0.4\times10^{20}$eV(just below the GZK cutoff) to
$1\times10^{20}$eV.
A comparison of anisotropies observed below and above the GZK cutoff
  will be a crucial test of whether clustering is important.
  For $r_0\sim 10$~Mpc, one can show that one will have to observe $10^2$
  ($10^3$ events) above $10^{20}$ ($4\times10^{19}$) eV to detect a
  $3\sigma$ effect. Thus, the detection of anisotropies would require the
  large exposure of the Auger project.

\noindent 4)
No characteristic dependence on Galactic coordinates is expected.
This is in contrast to many exotic particle-physics
scenarios for the production of ultra-high energy cosmic
rays in which a strong dependence on
Galactic coordinates is predicted,
in particular, a peaking toward the Galactic center. The Fly's Eye
 experiment reports an angular distribution peaked towards the Galactic
 disk for lower energy cosmic rays, which
 we interpret to be due to  a different, more local set of sources than the
 sources of ultra-high energy cosmic rays.

\noindent 5)
The suggested correlation of the directions of ultra-high energy
cosmic rays with the directions of compact radio-loud quasars (Farrar
\& Bierman 1998)
should disappear as more events become available. The candidates
suggested are all at distances $> 10^3$ Mpc, much  beyond the allowed
range of $\sim$ 40 Mpc for ultra-high energy protons.

\acknowledgments
We are grateful to an anonymous referee for constructive comments on
the presentation in the initial manuscript.
JNB is grateful to D. Eisenstein for instructive
conversations about correlations functions 
 and to NSF grant
\#PHY95-13835 for financial support.
Much of this work was done during a visit by JNB to the Weizmann
Institute of Science.

\newpage

\begin{figure}[h]
\centerline{\psfig{figure=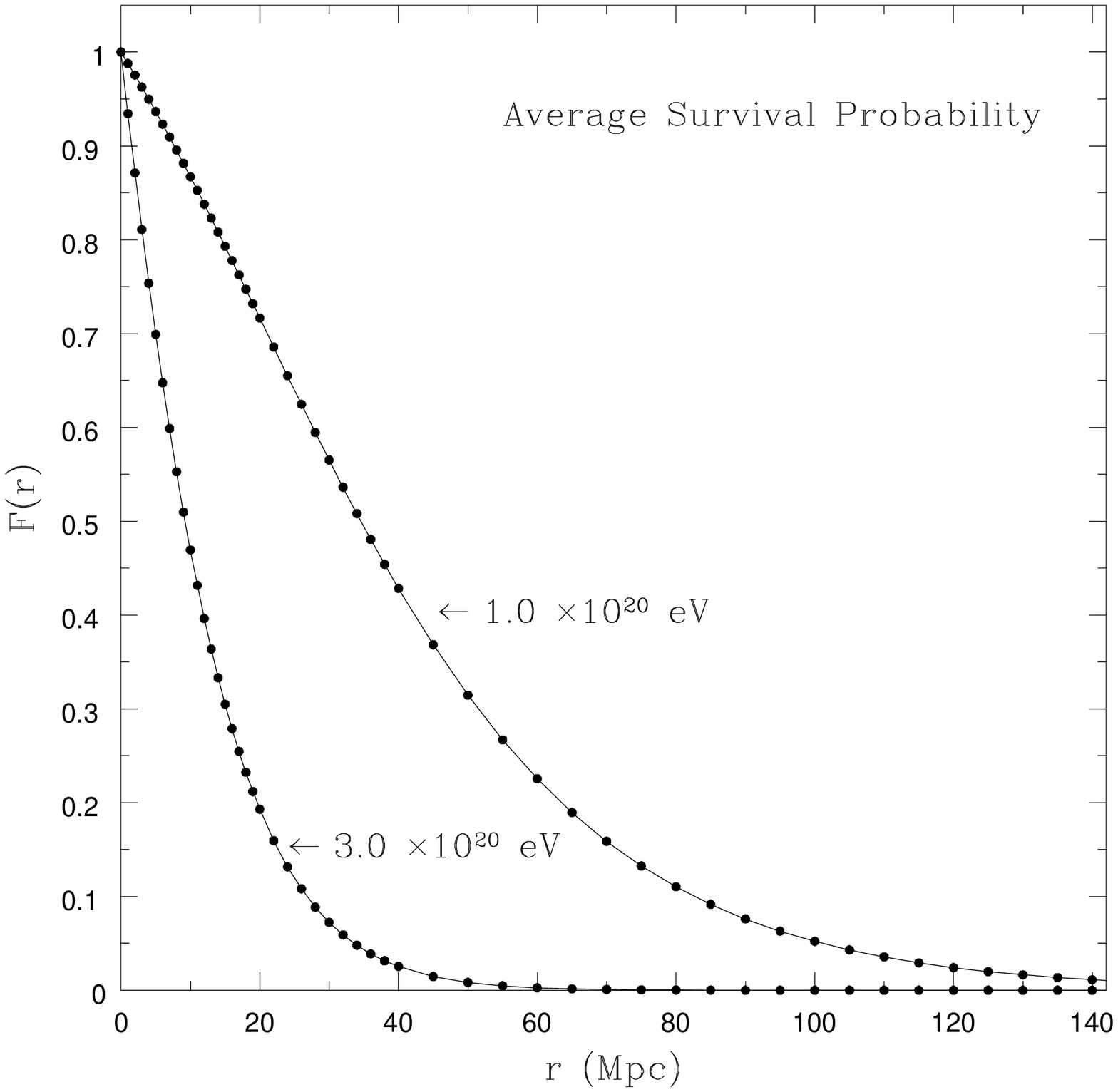,width=6in}}
\caption[]{The survival probability averaged over energies. The
plotted function, $F(r)$,  represents the  probability that a
proton created at a distance $r$ will reach the earth with an energy
in excess of the cutoff energy $E_c$. Results are shown for
$E_c = 1 \times 10^{20}$ eV and $E_c = 3 \times 10^{20}$ eV
with an energy spectrum $\partial S/\partial E \propto E^{-2}$.
}
\label{fig:function}
\end{figure}

\end{document}